%
% the following is to use blackboard bold fonts --
\let\useblackboard=\iftrue
%
% activate this if you don't have them.
%\let\useblackboard=\iffalse
%
% You might also need to remove this line.
\newfam\black
\input harvmac.tex
%\input tables.tex
%\input labeldefs.tmp
%
%%%%%%%%%%%%%%%%%%%%%%%%%%%%%%%%%%%%%%%%%%%%%%%%%%%%%%%%%%%%%%%
%The following lines are needed to insert the accompanying figures in
%the paper. If you do not have epsf, then comment out the line
% ``\input epsf'', and print the figures separately. The figures are
%at
%the end of the tex file, with instructions for their extraction.
%\input epsf.tex
\ifx\epsfbox\UnDeFiNeD\message{(NO epsf.tex, FIGURES WILL BE
IGNORED)}
\def\figin#1{\vskip2in}% blank space instead
\else\message{(FIGURES WILL BE INCLUDED)}\def\figin#1{#1}\fi
\def\ifig#1#2#3{\xdef#1{fig.~\the\figno}
\midinsert{\centerline{\figin{#3}}%
\smallskip\centerline{\vbox{\baselineskip12pt
\advance\hsize by -1truein\noindent{\bf Fig.~\the\figno:} #2}}
\bigskip}\endinsert\global\advance\figno by1}
%%%%%%%%%%%%%%%%%%%%%%%%%%%%%%%%%%%%%%%%%%%%%%%%%%%%%%%%%%%%%%%%
\noblackbox
\def\Title#1#2{\rightline{#1}
\ifx\answ\bigans\nopagenumbers\pageno0\vskip1in%
\baselineskip 15pt plus 1pt minus 1pt
\else%\special{papersize=11in,8.5in}%
\def\listrefs{\footatend\vskip
1in\immediate\closeout\rfile\writestoppt
\baselineskip=14pt\centerline{{\bf
References}}\bigskip{\frenchspacing%
\parindent=20pt\escapechar=` \input
refs.tmp\vfill\eject}\nonfrenchspacing}
\pageno1\vskip.8in\fi \centerline{\titlefont #2}\vskip .5in}
 
scaled\magstep3
 
scaled\magstep3
 
scaled\magstep3
 
scaled\magstep3
 
scaled\magstep3
\ifx\answ\bigans\def\tcbreak#1{}\else\def\tcbreak#1{\cr&{#1}}\fi

\useblackboard
\message{If you do not have msbm (blackboard bold) fonts,}
\message{change the option at the top of the tex file.}

\font\blackboard=msbm10 scaled \magstep1
\font\blackboards=msbm7
\font\blackboardss=msbm5
%\newfam\black
\textfont\black=\blackboard
\scriptfont\black=\blackboards
\scriptscriptfont\black=\blackboardss

\else

\fi
% *************************************
%\draftmode
%

%
\def\yboxit#1#2{\vbox{\hrule height #1 \hbox{\vrule width #1
\vbox{#2}\vrule width #1 }\hrule height #1 }}
\def\fillbox#1{\hbox to #1{\vbox to #1{\vfil}\hfil}}
\def\ybox{{\lower 1.3pt \yboxit{0.4pt}{\fillbox{8pt}}\hskip-0.2pt}}

\def\comments#1{}

\def\half{{1\over 2}}

\def\vev#1{\langle{#1}\rangle}

\def\CA{{\cal A}}

\def\II{\relax{I\kern-.07em I}}

\def\inbar{\,\vrule height1.5ex width.4pt depth0pt}
\def\IZ{\relax\ifmmode\mathchoice
{\hbox{\cmss Z\kern-.4em Z}}{\hbox{\cmss Z\kern-.4em Z}}
{\lower.9pt\hbox{\cmsss Z\kern-.4em Z}}
{\lower1.2pt\hbox{\cmsss Z\kern-.4em Z}}\else{\cmss Z\kern-.4em
Z}\fi}
\def\IB{\relax{\rm I\kern-.18em B}}
\def\IC{{\relax\hbox{$\inbar\kern-.3em{\rm C}$}}}
\def\ID{\relax{\rm I\kern-.18em D}}
\def\IE{\relax{\rm I\kern-.18em E}}
\def\IF{\relax{\rm I\kern-.18em F}}
\def\IG{\relax\hbox{$\inbar\kern-.3em{\rm G}$}}
\def\IGa{\relax\hbox{${\rm I}\kern-.18em\Gamma$}}
\def\IH{\relax{\rm I\kern-.18em H}}
\def\IK{\relax{\rm I\kern-.18em K}}
\def\IP{\relax{\rm I\kern-.18em P}}
\def\pp{{\relax{=\kern-.42em |\kern+.2em}}}
%\def\IX{\relax{\rm X\kern-.01em X}}
%this doesn't work

\font\cmss=cmss10 \font\cmsss=cmss10 at 7pt
\def\IR{\relax{\rm I\kern-.18em R}}

\def\frac#1#2{{{#1} \over {#2}}}

%
%Journal macros
%

\def\NP{{\it Nucl. Phys.\ }}

\def\PL{{\it Phys. Lett.\ }}
\def\PR{{\it Phys. Rev.\ }}
\def\PRL{{\it Phys. Rev. Lett.\ }}
\def\CMP{{\it Comm. Math. Phys.\ }}

\def\JHEP{{\it JHEP \ }}
\def\ATMP{{\it ATMP \ }}

\Title{ \vbox{\baselineskip12pt\hbox{hep-th/9903237}
\hbox{BROWN-HET-1176}\hbox{RH-02-99}\hbox{PUPT-1839}
}}
{\vbox{
\centerline{AdS/CFT and the Information Paradox}}}

\centerline{ David A. Lowe}
\medskip
\centerline{\it Department of Physics, Brown University, Providence,
RI 02912, USA}
\centerline{\tt lowe@het.brown.edu}
\medskip
\centerline{and}
\medskip
\centerline{L\'arus Thorlacius}
\medskip
\centerline{\it Department of Physics, Princeton University, Princeton,
NJ 08544, USA}
\centerline{\it and}
\centerline{\it University of Iceland, Science Institute, Dunhaga 3,
107 Reykjav\'{\char16}k, Iceland}
\centerline{\tt lth@raunvis.hi.is}
\bigskip

\centerline{\bf{Abstract}}

\noindent
The information paradox in the quantum evolution of black holes
is studied within the framework of the AdS/CFT correspondence.
The unitarity of the CFT strongly suggests that all information
about an initial state that forms a black hole is returned in
the Hawking radiation.  The CFT dynamics implies an information
retention time of order the black hole lifetime.  This fact
determines many qualitative properties of the non-local effects
that must show up in a semi-classical effective theory in the bulk.
We argue that no violations of causality are apparent to local
observers, but the semi-classical theory in the bulk duplicates
degrees of freedom inside and outside the event horizon.  Non-local
quantum effects are required to eliminate this redundancy.  This
leads to a breakdown of the usual classical-quantum correspondence
principle in Lorentzian black hole spacetimes.

\vfill

\Date{\sl March, 1999}

%References
\lref\hawkone{S.W. Hawking, ``Breakdown of Predictability in Gravitational
Collapse,'' \PR {\bf D14} (1976) 2460.}
\lref\hawktwo{S.W. Hawking, ``Particle Creation by Black Holes,'' \CMP {\bf 43}
(1975) 199.}
\lref\horog{G.T. Horowitz and H. Ooguri, ``Spectrum of Large $N$ Gauge
Theory from Supergravity,'' hep-th/9802116.}
\lref\graham{C.R. Graham and J.M. Lee, ``Einstein Metrics with
Prescribed Conformal Infinity on the Ball,'' {\it Adv. Math.} {\bf 87}
(1991) 186.}
\lref\witone{E. Witten, ``Anti De Sitter Space and Holography,''
\ATMP {\bf 2} (1998) 253, hep-th/9802150.}
\lref\luscher{M. Luscher and G. Mack, ``Global Conformal Invariance
and Quantum Field Theory,'' \CMP {\bf 41} (1975) 203.}
\lref\gubser{S.S. Gubser, I. Klebanov and A.M. Polyakov, ``Gauge
theory correlators from noncritical string theory,'' \PL {\bf B428}
(1998) 105, hep-th/9802109.}
\lref\wittwo{E. Witten, ``Anti-de Sitter Space, Thermal Phase
Transition, and Confinement in Gauge Theories,'' \ATMP {\bf 2} (1998)
505, hep-th/9803131.}
\lref\maldacena{J. Maldacena, ``The Large $N$ Limit of Superconformal
theories and gravity,'' \ATMP {\bf 2} (1998) 231, hep-th/9711200.}
\lref\itzhaki{N. Itzhaki, J.M. Maldacena, J. Sonnenschein and
S. Yankielowicz, ``Supergravity and the Large $N$ Limit of Theories
with Sixteen Supercharges,'' hep-th/9802042.}
\lref\mrbee{V. Balasubramanian, P. Kraus and A. Lawrence, ``Bulk
vs. Boundary Dynamics in Anti-de Sitter Spacetime,'' hep-th/9805171.}
\lref\mrbeetwo{V. Balasubramanian, P. Kraus, A. Lawrence and
S.P. Trivedi, ``Holographic Probes of Anti-de Sitter Spacetimes,''
hep-th/9808017.}
\lref\banks{T. Banks, M.R. Douglas, G.T. Horowitz and E. Martinec,
``AdS Dynamics from Conformal Field Theory,'' hep-th/9808016.}
\lref\susswit{L. Susskind and E. Witten, ``The Holographic Bound in
Anti-de Sitter Space,'' hep-th/9805114.}
\lref\susskind{L. Susskind, ``Strings, Black Holes and Lorentz
Contraction,'' hep-th/9308139.}
\lref\hologra{L. Susskind, ``The World is a Hologram,''
hep-th/9409089.}
\lref\marolf{G.T. Horowitz and D. Marolf, ``Where is the Information
Stored in Black Holes?'' hep-th/9610171.}
\lref\lowe{D.A. Lowe,
{``The Planckian Conspiracy: String Theory and the Black Hole
Information
Paradox,''} \NP {\bf  B456 } (1995) 257, hep-th/9505074.}
\lref\lpstu{
D.A. Lowe, J. Polchinski, L. Susskind, L. Thorlacius and J. Uglum,
{ ``Black Hole Complementarity vs. Locality,''} \PR {\bf
D52 } (1995) 6697, hep-th/9506138.}
\lref\green{T. Banks and M.B. Green, ``Non-perturbative effects in
$AdS_5 \times S^5$ string theory and $d=4$ Yang-Mills,''
\JHEP {\bf 5} (1998) 002, hep-th/9804170.}
\lref\sussthor{L. Susskind and L. Thorlacius, ``Gedanken Experiments
Involving Black Holes,'' \PR {\bf D49} (1994) 966, hep-th/9308100.}
\lref\page{D.N. Page, ``Average Entropy of a Subsystem,''
\PRL {\bf 71} (1993) 1291, gr-qc/9305007.}
\lref\pagetwo{D.N. Page, ``Information in Black Hole Radiation,''
\PRL {\bf 71} (1993) 3743, hep-th/9306083.}
\lref\horor{G.T. Horowitz and N. Itzhaki, ``Black Holes, Shock Waves,
and Causality in the AdS/CFT Correspondence,'' hep-th/9901012.}
\lref\das{S.R. Das, ``Brane Waves, Yang-Mills theories and
Causality,'' hep-th/9901004.}
\lref\rey{D. Bak and S.-J. Rey, ``Holographic view of causality and
locality via branes in AdS/CFT correspondence,'' hep-th/9902101.}
\lref\polchsuss{J. Polchinski and L. Susskind, ``Puzzles and Paradoxes
about Holography,'' hep-th/9902182.}
\lref\sussflat{L. Susskind,``Holography in the Flat Space Limit,''
hep-th/9901079.}
\lref\jpolch{J. Polchinski, ``S-Matrices from AdS Spacetime,''
hep-th/9901076.}
\lref\hawkpag{S.W. Hawking and D.N. Page, ``Thermodynamics of Black
Holes in Anti-de Sitter Space,'' \CMP {\bf 87} (1983) 577.}
\lref\lowetwo{D.A. Lowe, ``Statistical Origin of Black Hole Entropy in
Matrix Theory,'' \PRL {\bf 81} (1998) 256, hep-th/9802173.}
\lref\btzbh{M. Banados, C. Teitelboim and J. Zanelli, ``The Black Hole in
Three-Dimensional Spacetime,'' \PRL {\bf 69} (1992) 1849, hep-th/9204099.}
\lref\strominger{A. Strominger, ``Black Hole Entropy from Near-Horizon
Microstates,'' hep-th/9712251.}
\lref\peetross{A. Peet and S. Ross, ``Microcanonical Phases of String
Theory on $AdS_m\times S^n$,'' \JHEP {\bf 12} (1998) 020, hep-th/9810200.}
\lref\gregory{R. Gregory and R. Laflamme, ``Black Strings and p-Branes
are Unstable,'' \PRL {\bf 70} (1993) 2837, hep-th/9301052.}
\lref\wald{R.M. Wald, unpublished, 1993.}
\lref\wilc{P. Kraus and F. Wilczek, ``Self-Interaction Correction to Black
Hole Radiance,'' \NP {\bf B433} (1995) 403, gr-qc/9408003.}
\lref\reviews{For reviews of the information problem see f.ex.\ J. Preskill,
``Do Black Holes Destroy Information?'' hep-th/9209058;  S. Giddings,
``Quantum Mechanics of Black Holes,'' hep-th/9412138; A. Strominger,
``Les Houches Lectures on Black Holes,'' hep-th/9501071; L. Thorlacius,
``Black Hole Evolution,'' hep-th/9411020.}
\lref\joep{J. Polchinski, ``S-Matrices from AdS Space-Time,''
hep-th/9901076.}
\lref\vijay{V. Balasubramanian, S.B. Giddings, and A. Lawrence, ``What Do
CFT's Tell Us About Anti-de-Sitter Space-Times,'' hep-th/9902052.}
\lref\stu{L. Susskind, L. Thorlacius, and J. Uglum, ``The Stretched
Horizon and Black Hole Complementarity,'' \PR {\bf D48} (1993) 3743,
hep-th/9306069.}
\lref\horross{G. Horowitz and S. Ross, ``Possible Resolution of Black Hole
Singularities from Large N Gauge Theory,'' JHEP (1998) 9804:015,
hep-th/9803085.}
\lref\siddhart{S. Sen, ``Average Entropy of a Subsystem,'' \PRL {\bf
77} (1996) 1, hep-th/9601132.}

\newsec{Introduction}

Hawking's information paradox \hawkone\ is an important theoretical
problem, which must be resolved by any theory that claims to provide
a fundamental description of quantum gravity.  The usual argument for
information loss in black hole evolution is made in the context
of a low-energy effective theory, defined on a set of smooth
spacelike hypersurfaces in a geometry describing the formation and
subsequent evaporation of a large mass black hole.
The key assumption in the argument is that the effective theory is
a conventional local quantum field theory.  If, on the
other hand, we assume that black hole evolution is a unitary process
we are led to the conclusion that spacetime physics is non-local
on macroscopic length scales.  The nature of this non-locality must
be subtle, for it is certainly not apparent in our everyday low-energy
activities.  It should only become manifest under extreme kinematic
circumstances, such as those that relate inertial and fiducial
observers in a black hole geometry.

Some evidence for the required
sort of non-local behavior has been found in string theory \lpstu .
The commutator of operators corresponding to an observer inside the
event horizon and an observer, who measures low-energy Hawking
radiation well outside the black hole, is non-vanishing in string
theory, in spite of the fact that these observers are spacelike
separated.
These observers are related by a trans-Planckian boost,
but if one instead considers a pair of spacelike separated observers
with low-energy kinematics the effect is strongly suppressed and one
recovers conventional local causality.  These results suggest that
the usual reasoning for information loss fails in the context of
string theory but they have a limited range of validity, being
obtained by perturbative, off-shell calculations in light-front
string field theory. Similar arguments can also be made by convolving
appropriate wavepackets with the S-matrix \lowe. In this case a
macroscopic characteristic length scale appears in the amplitude,
indicating non-local effects. The analyticity of S-matrix is, however,
consistent with conventional causality.

Recent progress towards a non-perturbative formulation of string
theory has provided new tools with which to explore these issues.
In the present paper we re-examine the information
problem from a modern point of view, using in particular the AdS/CFT
correspondence \maldacena, which states that string theory in a certain
background spacetime is equivalent to a supersymmetric gauge theory
that lives on the boundary of the spacetime.  The fact that the
gauge theory is unitary strongly supports the view that no information
is lost in the quantum mechanical evolution of black holes, but it
is less clear how the unitarity is implemented from the spacetime
point of view.

In section~2 we briefly review the AdS/CFT correspondence in Euclidean
space, and discuss the definition of the Lorentzian correspondence by
straightforward analytic continuation. Black hole backgrounds, and
their CFT descriptions are described in section~3.  A simple gedanken
experiment is considered in section~4, which allows us to infer
qualitative features of the non-locality that must be present
in the effective theory in the bulk. The dual gauge theory description
implies an information retention time for the black hole, which plays
a crucial role in these arguments. We conclude that local observers
inside or outside the black hole see no violation of causality. We argue,
however, that the semi-classical effective theory duplicates degrees of
freedom inside and outside the event horizon, and that non-local quantum
effects are required to eliminate this redundancy. Classically there are
no such non-local effects, thus the usual classical-quantum correspondence
principle breaks down in black hole spacetimes.

\newsec{Review of AdS/CFT Correspondence}

As described in \refs{\witone, \gubser}, the natural relation
between CFT and AdS correlators is
\eqn\corresp{
\vev{ \exp \int_{S^d} \phi_0 \CO  }_{CFT} = Z(\phi_0)~,
}
where $\phi_0(\Omega)$ is the boundary value of a field in
the bulk, $\CO$ is the dual operator in the CFT,
and $Z$ is the string partition function on AdS space with
boundary conditions $\phi_0$. This statement is made in the Euclidean
formulation. It is made plausible by a remarkable theorem of Graham and Lee
\graham\ that says that for any sufficiently
smooth metric boundary values, there
exists a unique smooth solution in the bulk.

In the following we will need to formulate the AdS/CFT correspondence
in Lorentzian signature spacetime. Luscher and Mack \luscher\ have
shown that the Euclidean Green's functions of a quantum field theory
invariant under the Euclidean conformal group can be analytically
continued to Lorentzian signature, and the resulting Hilbert space of
states carries a unitary representation of the infinite-sheeted
universal covering group of the Lorentzian conformal group. The
natural Lorentzian spacetime is an infinite-sheeted
covering of Minkowski space and thus the natural spacetime to consider in
the bulk is the universal covering space of AdS.

Let us review how this analytic continuation proceeds for the case of
$AdS_5$.
Projective
coordinates for anti-de Sitter space $\xi^a$ with $a=1,\cdots 6$
satisfy
\eqn\projads{
(\xi^6)^2 - (\xi^4)^2 - (\xi^k)^2 = R^2~,
}
where $k=1,2,3,5$ and $R$ is the radius of curvature of the AdS space.
In the boundary limit we can drop the $R^2$ term, and parameterize the
coordinates as
\eqn\projch{
\xi^6 = r \cosh \sigma, \quad \xi^4 = r \sinh \sigma, \quad \xi^k = r e^k~,
}
where $e^k$ is a unit four-vector. The Euclidean conformal group
$SO(5,1)$ acts
in an obvious way on the $\xi$ coordinates.
The set of Euclidean
coordinates to be used for the boundary field theory are $(x^4, \vec
x)$ (with $\vec x = x^j$, $j=1,2,3$)
\eqn\euclcoor{
x^4 = {\sinh \sigma \over \cosh \sigma + e^5}, \quad \vec x = {\vec e
\over \cosh \sigma + e^5}~.
}
The analytic continuation corresponds to taking $\sigma = i \tau$ with
$-\infty < \tau < \infty$, which leads to an infinite-sheeted covering
of Minkowski space $\tilde M$ with coordinates $(\tau, e^k)$.

A single copy of Minkowski space can be embedded in $\tilde M$ by
taking the subspace $-\pi<\tau < \pi$ and $e^5 > - \cos \tau$. The
usual Minkowski coordinates are
\eqn\minkco{
x^0 = {\sin \tau \over \cos \tau +e^5}, \quad \vec x = {\vec e \over
\cos \tau + e^5}~.
}
This space is conformal to the boundary at infinity of the Poincare
patch of anti-de Sitter space, which is defined by the coordinates $(x^\mu,z)$
\eqn\poinpat{
ds^2 = {R^2 \over z^2} ( dx^2 -dz^2)~.
}
The boundary at infinity corresponds to $z=0$. Scale/radius duality is
manifest in this set of coordinates as illustrated by the
D-instanton/Yang-Mills instanton duality. The scale factor of the
Yang-Mills instanton translates into the position along the radial $z$
coordinate of the D-instanton in the bulk \green.

To generate translations with respect
to the global time $\tau$, one acts with the conformal Hamiltonian of
the field theory $H= \half(P^0 + K^0)$. Luscher and Mack show $H$ is
positive and self-adjoint \luscher, and that there is a unique vacuum
state annihilated by $H$, invariant under conformal transformations.

In the following, we will take the point of
view that the Lorentzian theory in the bulk is defined by this
analytic continuation of the Euclidean correlation
functions. As we will see later, this is a rather subtle point.
Alternative proposals for the Lorentzian version of the
AdS/CFT correspondence have appeared in the literature. An example is
the eternal black hole solution considered in \mrbeetwo, where the
dynamics is described instead by two disconnected boundary field theories.

\newsec{The Information Puzzle and AdS black Holes}

Hawking's information paradox arises when one considers the quantum
mechanical evolution of black holes \refs{\hawkone,\hawktwo}.
The issues are most sharply
defined for a black hole that is formed by gravitational collapse
from non-singular initial data and subsequently evaporates by emitting
apparently thermal Hawking radiation.  If the initial configuration
is described by a pure quantum state and the Hawking radiation is
truly thermal then this process involves evolution from a pure state
to a mixed one, which violates quantum mechanical unitarity.
A related problem involves perturbations on a background extremal black
hole.  The resulting non-extremal black hole will emit Hawking radiation
until it approaches extremality once more and the question of
unitarity arises in this context.  In both of the above
settings one can also consider an equilibrium configuration
where energy is fed into a black hole at the same rate that it
evaporates.  In this case the paradox arises from the fact that
an arbitrary amount of information can be encoded into the
infalling matter over time and most of this information will be
absent from the outgoing Hawking radiation if it is truly thermal.

In order to take advantage of some of the recent developments in
fundamental gravitational theory one would like to formulate analogous
questions in the context of black hole evolution in anti-de Sitter
spacetime.  This presents us with some immediate problems.  For one
thing AdS gravity does not have a well posed initial value problem
due to the global causal structure of the AdS spacetime.  While
spacelike slices of AdS spacetime have infinite volume, null signals
can propagate from infinity into any given region in a finite affine
parameter.  As a result it is problematic to define unitary quantum
mechanical evolution in AdS space even in the absence of black holes.

This problem can be circumvented in a number of ways.  We can for
example impose boundary conditions at infinity, which in the absence
of black hole formation lead to unitary evolution.  The important point
is that within some such framework we can study the formation and
evaporation of a black hole whose lifetime is short compared to the
light-crossing time of the AdS geometry.  We can choose parameters in
such a way that this black hole is nevertheless large compared to the
Planck scale and thus carries a significant amount of information.
The question of possible information loss associated with the evolution
of the black hole is then effectively decoupled from the unitarity
problem of the underlying AdS geometry.

There also exist black
hole solutions in AdS space where the Schwarzschild radius is large
compared to the characteristic AdS length scale.  For such black holes
there is no separation of scales and thus
difficult to disentangle the two unitarity problems.  On the other
hand, as we shall see below, such black holes are extremely
unstable and not so useful for studying the information problem
in the first place.

Another way to proceed would be to consider the asymptotically flat
geometry of an extremal D-brane, whose near-horizon limit is locally
isomorphic to AdS spacetime.  The asymptotically flat region then
regulates the infrared pathology of the AdS geometry and one can
consider evolution from initial data (subject to physical boundary
conditions at the D-brane horizon).  One would again want to study black
holes which are small compared to the characteristic AdS length scale
which in this case is the size of the extremal D-brane throat.
There of course also exist solutions describing black holes that
are larger than the throat scale.  These are just the non-extremal
$p$-branes of the higher-dimensional supergravity theory.  They have
non-vanishing Hawking temperature and their evolution leads to an
information problem of the usual type.  On the other hand, once we
are far from extremality the gauge theory correspondence, which is
the main new tool at our disposal, is no longer useful.  There is
a rather subtle problem with regulating the infrared problems in this
manner.  The thermodynamic behavior is sensitive to the asymptotic
boundary conditions, {\it i.e.} whether the conformal boundary in
Euclidean signature is taken to be $S^n\times S^1$ or
${\bf R}^n\times S^1$ \peetross .  In the following, we make the
choice $S^n\times S^1$, which is appropriate for the black holes we
want to study, whereas the near-horizon limit of a D-brane gives rise
to ${\bf R}^n\times S^1$.

The metric of a static Schwarzschild black hole in $n+1$-dimensional
asymptotically AdS spacetime can be written
\eqn\schw{
ds^2 = -\bigl({r^2\over R^2} +1 - {\mu\over r^{n-2}}\bigr) dt^2
+\bigl({r^2\over R^2} +1 - {\mu\over r^{n-2}}\bigr)^{-1} dr^2
+r^2 d\Omega_{n-1}^2 \,,
}
where $R$ is the AdS radius of curvature and $\mu$ is proportional
to Newton's constant in $n+1$ spacetime dimensions times
the black hole mass,
\eqn\mass{
\mu = {8 \Gamma({n\over 2}) G_N M\over (n-1) \pi^{(n-2)/2}} \,.
}
As we approach the black hole from large $r$ the metric has a
coordinate singularity at the AdS-Schwarzschild radius $r=r_s$, where
\eqn\schwrad{
{r_s^2\over R^2}+1-{\mu\over r_s^{n-2}} = 0 \,.
}
In the limit of small black hole mass, $\mu << R^{n-2}$, the black
hole parameters approach those of a black hole of equal mass in
asymptotically flat spacetime,
\eqn\rssmall{
r_s \approx \mu^{1/(n-2)} \,,
}
while in the large mass limit, $\mu >> R^{n-2}$, we instead have
\eqn\rslarge{
r_s \approx (\mu R^2)^{1/n} \,.
}
One obtains the Hawking temperature in the standard way by
continuing to the Euclidean section and requiring the horizon to
be smooth,
\eqn\thawking{
T_h = {nr_s^2 +(n-2)R^2 \over 4\pi R^2 r_s} \,.
}
In the small mass limit this reduces to $T_h \approx (n-2)/4\pi r_s$,
which is the usual Hawking temperature of a Schwarzschild black hole
in asymptotically flat spacetime, but in the large mass limit we find
that the AdS black hole has positive specific heat,
$T_h \approx nr_s/4\pi R^2$.

We can now use the Stefan-Boltzmann law to estimate the lifetime of
an AdS black hole.  For small black holes we find
\eqn\evapsmall{
{d\mu\over dt} \sim \mu^{-2/(n-2)} \,,
}
leading to a lifetime which grows as a power of the black hole mass,
\eqn\lifesmall{
\tau \sim \mu^{n/(n-2)} \,.
}
If we choose parameters so that
$r_s$ is a macroscopic length (in an AdS background with even larger
radius of curvature) then the black hole will slowly evaporate, at
a rate reliably predicted by semi-classical considerations.

In the large mass limit the behavior is very different.  In this
case the evaporation rate grows with mass,
\eqn\evaplarge{
{d\mu\over dt} \sim \mu^2 \,,
}
leading to a lifetime that is bounded from above\foot{Here we assume
the boundary conditions at infinity correspond to zero incoming
flux. For reflecting boundary conditions instead, the black hole will rapidly
come into equilibrium with the thermal radiation. We thank G. Horowitz 
for discussions on this point.}
\eqn\lifet{
\tau_0 -\tau \sim 1/\mu \,.
}
The approximation of large mass breaks down as $r_s\rightarrow R$,
so $\tau$ should be interpreted as the time that elapses before
the black hole has evaporated to a size of order the AdS length
scale.  The subsequent evaporation rate will be independent of the
original black hole mass so the total lifetime is obtained by adding
some constant to $\tau$.  The parameter $\tau_0$ in equation
\lifet\ is the value of $\tau$ in the limit where the original
black hole mass becomes infinite.  A black hole of arbitrarily large
initial mass will reduce to a size of order the AdS scale within
this time, which means that such objects are violently unstable.
They do not provide us with the slowly evolving background geometries
that are required for setting up the information puzzle.  In fact the
instability will most likely prevent them from forming in gravitational
collapse in the first place.  This does not preclude their
existence in thermal equilibrium with a high-temperature thermal bath
but since the heat bath is already in a mixed quantum state
that is not the ideal configuration for studying the information problem.

The upshot of all this is that, for the purpose of studying information
issues in black hole evolution, we want to consider black holes that
are macroscopic, {\it i.e.} large compared to the string scale, but
at the same time small compared to the AdS scale.  This means that their
Schwarzschild radius is also small compared to the radius of the
transverse compact space that accompanies the AdS geometry.  The favored
configuration in this case is in fact not the AdS/Schwarzschild black hole
that we have been discussing but rather a higher-dimensional black hole
that is localized somewhere on the transverse compact space \gregory .
This is not really a problem for our discussion.  The semi-classical
expressions \rssmall\ for the Schwarzschild radius, \evapsmall\ for the
evaporation rate, and \lifesmall\ for the black hole lifetime still remain
valid if we remember to replace the $n$ of $AdS_n$ by the number of space
dimensions of the higher-dimensional geometry.

The $2+1$-dimensional black hole is a rather special case.
The metric for the non-rotating BTZ black hole \btzbh\ takes the form
\eqn\btzmet{
ds^2 = ({r^2 \over R^2} - m) dt^2 - ({r^2 \over R^2} - m)^{-1} dr^2 -
r^2 d\varphi^2~,
}
where $\varphi$ has period $2\pi$.  The relationship between $m$ and
the black hole mass depends on which geometry one uses as a zero-mass
reference.  Two choices offer themselves.  One is to define $AdS_3$ to
have zero mass, in which case we have
\eqn\adsmass{
m = -1 + 8 G_N M_{adS} \,.
}
Since $m$ is required to be positive we see that $2+1$-dimensional black
holes have a non-vanishing minimum mass with this definition.  The other
definition, which is perhaps more natural from the point of view of black
hole physics, is to define the $m=0$ geometry in \btzmet\ to have vanishing
mass, so that
\eqn\btzmass{
m = 8 G_N M_{BTZ} \,,
}
In this case the Schwarzschild radius, $r_s = \sqrt{m} R$, goes to zero
when the mass is taken to zero, and $AdS_3$ appears as an isolated smooth
geometry in a family of solutions with naked singularities which formally
have negative mass.

The Hawking temperature of the black hole \btzmet\
is $T_h=\sqrt{m}/2\pi R$ and the entropy is
$S=\pi \sqrt{m} R/2 G_N$.  The lifetime of such $2+1$-dimensional black
holes is formally infinite.  This is because the rate of Hawking radiation
slows down as $m$ approaches zero.  This is not a problem because such
small $2+1$-dimensional black holes are not relevant to the physics.
Here the AdS/CFT correspondence involves string theory on a background of
the form $AdS_3\times S^3 \times M$ and black holes with Schwarzschild
radius small compared to the size of the $S^3$ are unstable to form
$5+1$-dimensional black holes that are localized on the $S^3$.
Those black holes have a finite lifetime, given by \lifesmall\ with $n=5$.

Let us now consider the description of macroscopic black holes, that are
nevertheless small compared to the scale of the transverse geometry,
from the gauge theory point of view. We begin with the $AdS_5$ case.
In the canonical ensemble, Witten \wittwo\ has shown there is a phase
transition from a large mass AdS-Schwarzschild solution to an AdS
space with certain discrete identifications,
generalizing the work of Hawking and Page
for $3+1$ dimensions \hawkpag. In the gauge theory, this
is reflected as a deconfinement transition in the gauge theory as the
temperature is lowered. For the $AdS_3$ case, there is no analog of
the Hawking-Page phase transition, and instead there is
a smooth cross-over as the temperature decreases.

To obtain a stable phase containing the intermediate
mass black holes that we will be interested in, it is convenient to
consider the microcanonical ensemble instead. In general it will be
necessary to impose additional constraints on the ensemble, by
requiring that the energy density be sufficiently well-localized, to
ensure that only single black hole states dominate the ensemble.
The gauge theory version of this ensemble will likewise be a
microcanonical ensemble with additional constraints. To determine the
form of these constraints one must follow through the mapping of the
energy-momentum tensor of the gravity theory into operators in the
gauge theory. This is of course a difficult task once one wishes to go
beyond the linearized approximation, but is nevertheless a
well-defined procedure.
Analogous constraints appear in the discussion of black
hole entropy in Matrix theory given in \lowetwo.

\newsec{Unitarity vs.\ Locality}

The static Schwarzschild solution \schw\ describes a black hole in
equilibrium with a thermal gas in an AdS background.  In order to
study the information problem we want instead to consider the evolving
geometry of a black hole which forms by gravitational collapse in AdS
space and then evaporates as it emits Hawking radiation.  We will not
attempt to write down such a solution but rather choose parameters in
such a way that the black hole evaporates slowly compared to all
microscopic timescales and has a long lifetime.  The geometry is then
described to a good approximation by \schw\ with $\mu$ varying slowly
with time.  In order to separate the issue of black hole information
loss from the usual unitarity problem in AdS space we will also assume
that this macroscopic black hole is formed in an AdS background with a
very large radius of curvature so that the lifetime  \lifesmall\ is
small compared to the light-crossing time.

We can now imagine describing the bulk of the evolution of the black
hole in terms of a low-energy effective theory defined on a set of
`nice' slices \refs{\wald,\wilc,\lpstu} that foliate the slowly evolving
spacetime and approach the local free-fall frame of infalling matter at
(and inside) the event horizon but also approach the frame of fiducial
observers far away from the black hole.  If the low-energy effective theory
on the nice slices is a local quantum field theory then it follows from
standard arguments \hawktwo\ that the quantum state of the Hawking
radiation will be correlated to that of the infalling matter.  If we
further assume that the black hole completely evaporates, leaving only
outgoing Hawking radiation behind, then the final state cannot be a pure
quantum state.  On the other hand, the evolution of states in a local
quantum theory is unitary and therefore the final state should be a
pure state if the initial configuration before the black hole forms is
described by a pure state.

This apparent contradiction must somehow be resolved in a fundamental
theory and a number of scenarios have been put forward \reviews . The
conjectured AdS/CFT correspondence supports the view that black hole
evolution is unitary since the gauge theory is manifestly unitary.
This in turn means that the low-energy effective theory cannot be a
local quantum field theory\foot{A possible loophole to this argument
would be that information is stored in Planck-mass remnant states.
The existence of such remnants would imply an enormous peak in the
density of states around the Planck scale.  This, however,
is in conflict with the gauge theory calculations of black hole
entropy.}.  This is not a problem in the AdS/CFT
context because the duality map that relates the gauge theory and
spacetime physics is quite non-local, as has been emphasized in
recent work \refs{\sussflat,\joep,\vijay}.

In the following we will assume the validity of the AdS/CFT conjecture
and ask what its implications are for the propagation of information
in black hole spacetimes.  For this purpose it is useful to consider
gedanken experiments which highlight the conflict between unitarity
and locality \sussthor .  Let us in particular examine a simple
experiment which involves correlated degrees of freedom inside and
outside the event horizon.  Imagine a pair of spins prepared in a
singlet state well outside the black hole.  Here `spin' should be
understood as some internal label because conventional spin can in
principle be detected by its long range gravitational field and is
therefore not suitable for this experiment.   One of the spins is then
carried inside the black hole, where a measurement of the spin is made,
at some point $\CA$.  Meanwhile, an observer $\CO$ outside the black
hole makes measurements on the Hawking radiation. If all the information
about the quantum state inside the black hole is encoded in the Hawking
radiation, this observer can effectively measure a component of the
spin that went inside the black hole. The observer then passes inside
the event horizon, where he can receive a signal from point $\CA$,
which potentially contradicts his previous measurement, in
violation of the laws of quantum mechanics.
There is no real paradox here from the spacetime point of view,
for if $\CO$ is to learn of the contradiction before hitting the
singularity then the signal sent by $\CA$ has to involve frequencies
beyond the Planck scale \sussthor .  If, on the other hand, the signal
from $\CA$ is generated using only low-energy physics then $\CO$ will
have entered a region of strong curvature before receiving it.
Either way the analysis of the gedanken experiment requires knowledge
of physics beyond the Planck scale and the apparent contradiction only
arises if we make the (unwarranted) assumption that this physics is
described by local quantum field theory.

Let us reconsider this experiment using the boundary gauge theory.
There is only a single quantum state on any given time slice from the
boundary point of view, so no contradiction can arise between the
spin measurements.  During the evaporation phase, the Hamiltonian
of the boundary theory, to a good approximation, generates evolution
in the asymptotic time $t$ in the AdS-Schwarzschild spacetime
\schw .  This time variable belongs to a coordinate system which
only covers a region of spacetime exterior to the black hole and
is therefore awkward for describing the fate of observers that enter
the black hole.  The history of such an observer is more economically
described if we instead evolve our quantum state using a different
timelike generator, one which is associated with the free fall frame
at the horizon and connects the interior and exterior regions of the
black hole \banks .  It is, however, an important matter of principle
that the evolution in asymptotic time  must contain all information
about the infalling matter, even inside the black hole region.  This
is guaranteed by the unitarity of the boundary gauge theory.

The scale/radius duality \maldacena\ tells us that an object far
outside the black hole is described by a localized configuration in
the gauge theory, but as the black hole is approached the same object
will be represented by an excitation of much larger transverse size
in the gauge theory if the system evolves in asymptotic time.
This should be a correct approximate statement in asymptotically
AdS backgrounds, but note that here we are going beyond the
application of scale/radius duality in the unperturbed AdS background.
The black hole is represented by an system of particles in the gauge
theory with fixed total energy. As an object
falls into the black hole the gauge theory configuration that
describes it spreads in transverse size and
at the same time gets entangled with the particles that make up
the black hole.
While the exact quantum state of the system of
object plus black hole contains the information that the object
continues its plunge towards the singularity, this information is
not readily available to outside observers.  In fact the only way
to access it is through careful observations of correlations in
the entire train of outgoing Hawking radiation, as we discuss further below.

Let us return to our gedanken experiment.  In order for an outside
observer to conduct a measurement on the spin that entered the black
hole, he or she measures correlations between  Hawking particles
emerging from the black hole at different times.   One could imagine
a more active type of measurement where the outside observer
attempts to probe the black hole in various ways.  This would only
serve to excite the black hole and be counterproductive since the
state of the spin would now be entangled with that of the probes in
addition to the original black hole.

The gauge theory configuration that describes the black hole
containing the spin at $\CA$ behaves as a conventional thermodynamic
system. Ideally we would like to calculate the entropy and lifetime
using the CFT description, but such a calculation requires a better
understanding of the strongly coupled CFT in the large $N$ limit.
But we stress these are nevertheless completely well-defined
computations in the CFT. The best we can do at present is to
assume the validity of the AdS/CFT
correspondence and infer that the gauge theory answers
coincide with the semi-classical gravity results.

With an understanding that the entropy and the rate of Hawking radiation
can be obtained from the CFT point of view, we can then invoke a result
of Page \refs{\page,\siddhart}. This states
that no useful information is emitted from a thermodynamic
system that is radiating, until its
coarse grained entropy has been reduced by a factor of two.  The time
this takes for an evaporating black hole is of order the black hole
lifetime.
In other words, there is an information retention time
in the gauge theory description of the black hole.  The
existence of an information retention time was postulated in
references \sussthor\ and \pagetwo\ but the AdS/CFT correspondence now
provides a concrete realization of that idea.

The information transfer between inside the event horizon and outside
the event horizon thus effectively takes place only when a time of order
the lifetime of the black hole has elapsed, from the point of view of a
distant outside observer.  It is also important to determine at what
point the outside measurements begin to have significant influence on
infalling observers inside the black hole.  Let us think of this, for
the moment, in the context of a low-energy effective theory, defined
on a set of `nice' slices.  If we assume that the effective theory
is a local field theory that has been evolved forward from a
non-singular initial configuration described by some pure quantum
state, and we further stipulate that all information about the initial
state is to be found in the outgoing Hawking radiation, then the
degrees of freedom on that part of the nice slice that is inside
the black hole can carry no information about the initial state \stu .
In other words, all information about the initial state must be
`bleached' out of the infalling matter immediately upon crossing
the event horizon, in blatant violation of the equivalence principle.
For a large mass black hole the horizon sits in a region of weak
curvature where tidal effects are small and an object in free fall
should pass through more or less unaffected.  This is, of course,
just a statement of the information paradox in the context of a
local effective field theory of gravity.

In the boundary theory this appears in the form of a somewhat
different puzzle.  There the infalling object is described by a
spreading field configuration which is getting entangled with the
ambient fields describing the black hole.  This entanglement gets
very complicated as the object is `thermalized' into the black hole
configuration, yet it is very delicate.  The slightest change in
the combined configuration could drastically change the results of
subsequent correlation measurements on the Hawking radiation,
leading the outside observers to conclude that the infalling
object did indeed get bleached as it passed through the horizon.

For a resolution of this puzzle we again have to appeal to the
AdS/CFT correspondence.  Since we know that an infalling object
encounters no obstacle at the horizon of a large black hole in
the supergravity description, the gauge theory dynamics must
somehow miraculously preserve the integrity of infalling matter
even if it appears, to a casual outside observer, to be
thermalized as it interacts with the black hole configuration.
This is very reminiscent of the discussion in reference \sussflat\
of low-energy, large impact parameter gravitational scattering.
On the supergravity side the particles hardly interact at all and
move past each other with only a small deflection, but on the gauge
theory side the objects completely merge during the collision and
interact strongly, but then somehow disentangle themselves and go
their separate ways.

Some recent work \refs{\vijay,\das\horor{--}\rey} 
has shown explicitly in certain
examples how the gauge theory
dynamics, at large N, leads to a local and causal
description of semi-classical low-energy supergravity.
Building on this, we can
argue that the
information transfer to the outgoing Hawking radiation must take
place near the black hole singularity from the point of view of
infalling observers.  Consider once again the gedanken experiment
involving spins.  The observer inside the black hole cannot be
influenced by the outside measurements until their result has been
communicated inside, but the outside measurements cannot be
completed until at least of order the black hole information
retention time has passed according to asymptotic AdS-Schwarzschild
clocks.  In order to receive a signal carrying outside results
before hitting the singularity, the observer inside would have to
undergo a proper acceleration of order $\exp(M^{2/(n-2)})$ (in $n+1$
dimensions). Such an acceleration would require trans-Planckian
energies, which are simply not available to a low-energy
observer.
We therefore conclude that the observation of Hawking radiation
does in fact bleach out a low-energy observer inside the horizon,
but the bleaching only takes place at the singularity, where life
is less than good anyway.

Since the singularity is by definition not in the causal past of
any of the outgoing Hawking radiation it is clear that this
information transfer is non-local on a macroscopic scale.
Causality in the boundary gauge theory does not guarantee causality
on the spacetime side.  It does lead to an approximate causality in
the bulk physics in flat or near-flat spacetime, but
our black hole example illustrates that even this macroscopic
causality must break down near spacetime singularities.
We note that this breakdown occurs already at the level of a
perturbative $1/N$ expansion in the boundary gauge theory.
The time evolution operator must be unitary, order by order in a $1/N$
expansion.

Further insight into the nature of this non-locality in the nice slice
theory can be gained by considering physics in the Euclidean continuation.
The analytic continuation from Euclidean space
completely determines arbitrary correlation functions in the
Lorentzian CFT, and in particular correlators
corresponding to the above gedanken experiment.
We learn from the theorem of Graham and Lee \graham\ that in Euclidean
signature, the classical bulk geometry is smooth, and
there is a one-to-one mapping between the field configurations in the
bulk, and those on the boundary. In Euclidean signature, we see no
sign of any non-locality in the bulk theory.
Only when we analytically continue
to Lorentz signature does the singularity arise,
hidden behind the event horizon. Classically, this leads to
too many degrees of freedom in the bulk. The degrees of freedom
inside and outside the
event horizon are independent. Quantum mechanically, the picture in
Lorentz signature is radically different. The CFT tells us the degrees
of freedom inside are to be identified with degrees of freedom outside
the event horizon.  This implies the usual classical-quantum
correspondence principle breaks down for black hole spacetimes.
The degrees of freedom in the correct quantum description in the bulk
do not smoothly go over to the classical degrees of freedom of the
supergravity theory.  The information paradox arises when we ask
questions involving these degrees of freedom that are duplicated in
the classical theory.  Non-local effects are
required in a semi-classical description on a set of nice slices,
to see that these degrees of freedom are in fact redundant.

This set of arguments also implies the CFT formulation resolves the
singularity of the black hole \refs{\horross,\banks} and allows us to
propagate states smoothly past the point where the black hole has
evaporated. Classically, this point looks like a timelike (or null)
naked singularity in the spacetime. Thus from the Lorentz point of view,
the analytic continuation from Euclidean signature leads to definite
boundary conditions on this naked singularity.

\bigskip
{\bf Acknowledgments}

The research of D.L. is supported in part by DOE grant
DE-FE0291ER40688-Task A.  The research of L.T. is supported
in part by a DOE Outstanding Junior Investigator award
DE-FG02-91ER40671.

\listrefs
\end